\documentclass[journal]{IEEEtran}

\usepackage{graphicx}

\usepackage{amsmath}
\ifCLASSOPTIONcompsoc
 \usepackage[caption=false,font=normalsize,labelfont=sf,textfont=sf]{subfig}
\else
 \usepackage[caption=false,font=footnotesize]{subfig}
\fi
\begin{document}
%
\title{Comparison of Experimental and Theoretical Mechanical Jitter in a THz Communication Link}
%
%
%

\author{Ethan~Abele,~\IEEEmembership{Graduate Student Member,~IEEE,}
Karl Strecker,
John F. O'Hara,~\IEEEmembership{Senior Member, IEEE}%
\thanks{Funding support: This work was supported in part by the National Aeronautics and Space Administration, Grant 80NSSC22K0878, and the National Science Foundation, Grant ECCS-2238132.}%
\thanks{Ethan Abele, Karl Strecker, and John F. O'Hara are with the School of Electrical and Computer Engineering, Oklahoma State University, Oklahoma, USA (e-mail: eabele, karl.l.strecker, oharaj{@okstate.edu})}%
}

\maketitle


\begin{abstract}
The effect of mechanical vibration (jitter) is an increasingly important parameter for next-generation, long-distance wireless communication links and the channel models used for their engineering. Existing investigations of jitter effects on the terahertz (THz) backhaul channel are theoretical and derived primarily from free space optical models. These lack an empirical and validated treatment of the true statistical nature of antenna motion. We present novel experimental data which reveals that the statistical nature of  mechanical jitter in 6G links is more complex than previously assumed. An unexpected multimodal distribution is discovered, which cannot be fit with the commonly cited model. These results compel the refinement of THz channel models under jitter and the resulting  system performance metrics. 
\end{abstract}


%
\IEEEpeerreviewmaketitle

\section{Introduction}

\IEEEPARstart{A}{ccurate} statistical modeling of microvibration and platform jitter is a highly important and growing concern in THz communication systems. Jitter effects are a consequence of the high antenna gain required to overcome free space path loss (FSPL). By definition, the beamwidths of these high gain antennas are extremely narrow. This allows unintended motions of the transmitter or receiver, referred to here as mechanical jitter, to cause significant signal fluctuations.

Mechanical jitter is typically considered to be a random process that must be modeled statistically. The probability distribution function (PDF) of the misalignment gain, $h_m$, is commonly used. This random variable quantifies the signal degradation as a function of antenna misalignment. Existing analytical models for free-space optical (FSO) links have recently been applied to THz communication systems \cite{abdalla_performance_2023,boulogeorgos_analytical_2019,boulogeorgos_outage_2020,boulogeorgos_joint_2022}. These works assume a Gaussian distribution of misalignment angles (azimuth and elevation) to derive an analytical PDF of $h_m$.  Since THz and FSO systems have very different physical antenna structures, the accuracy of this assumption becomes questionable.  This creates uncertainty in performance estimates such as bit error rate (BER) and outage probability. Experimental validation is a critical next step.

In this work we describe the first experimental measurements (to our knowledge) of the  distribution of misalignment gain in a long-distance THz wireless communication system. It was found that the experimental measurements do not match the existing (Gaussian) theoretical models, but show a rich and complex behavior indicative of multi-modal vibrations within the antenna structure.  The key differences are explained, and a causal mechanism is proposed.

 

\section{Measurement Procedure}

\subsection{Experimental Setup}

\begin{figure}[h]
\centering
\includegraphics[width=1.0\linewidth]{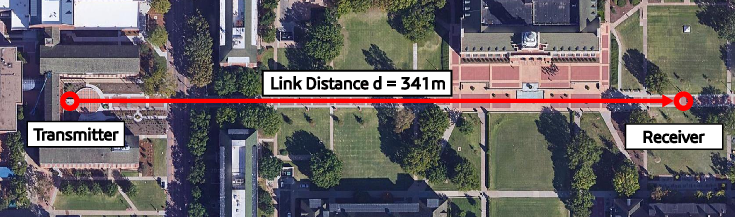}
\caption{Aerial view of measurement path.}
\label{fig:measurement-path}
\end{figure}

The THz communication system is well established and has been used to achieve  $>8$ gigabit/second communication over a 2.9~km range \cite{strecker2023wideband,o2024long}. Here, it is configured to generate a simple carrier tone at $f_c=130$~GHz, which is transmitted between a pair of equivalent, high gain, Cassegrain reflector antennas. The carrier was transmitted along an open line of sight (LOS) path of approximately 341~m on the campus of Oklahoma State University. The path is shown in Fig.~\ref{fig:measurement-path}. It consists of a flat and wide (6.8~m) walking path which is free from major obstructions within the expected beam radius.
\vspace{-0.2cm}

\vspace{-0.2cm}

The transmitter and receiver use identical cassegrain antennas with an approximate directivity of 50 dBi and a half power beamwidth of  $0.5^{\circ}$. The first null beamwidth is $0.7^{\circ}$. Such a narrow beamwidth is a necessary tradeoff to achieve a usable signal-to-noise ratio, but it clearly illustrates the importance of quantifying jitter or vibration effects. Even a small angular motion of either antenna can lead to significant degradation of received signal strength. 

In this work we are interested capturing the \textit{statistics} of the misalignment gain. In other words, we seek to measure the fluctuation in received power under jitter and observe its distribution. The studied jitter effects vary on the order of tens of hertz. Therefore, long duration measurements (several seconds) with high-fidelity sampling of the carrier are emphasized. To achieve this, the received signal is mixed down to the intermediate frequency (IF) of only $f_{IF}=400$~kHz. This would be too low for a practical communications waveform, but it is suitable for exploring the misalignment gain statistics since we are only interested in the amplitude fluctuations caused by random antenna motion. The downconverted signal is sampled at 5~MSps (megasample/second), more than $10\times$ the IF frequency. This configuration allows fine fluctuations of the received signal amplitude to be faithfully recorded and enables up to 10 seconds of continuous sampling before the memory depth of the recording oscilloscope is exceeded. 

Controlled jitter was applied to the system through the use of an unbalanced metal weight gripped in the chuck of an electric drill. This assembly was firmly attached to the instrument cart on which the transmitter and its antenna structure were mounted. The intensity of jitter can be roughly controlled by altering the revolutions per second (RPS) of the unbalanced weight, with a higher RPS producing more intense jitter at the antenna base. We define the following categories in this work: Baseline = no vibration applied, Low Jitter = 6.8 RPS motor speed, Moderate Jitter = 7.9 RPS motor speed, High Jitter = 10.3 RPS motor speed. 

\begin{figure*}[!t]
\centering
\subfloat[]{\includegraphics[width=0.3\linewidth]{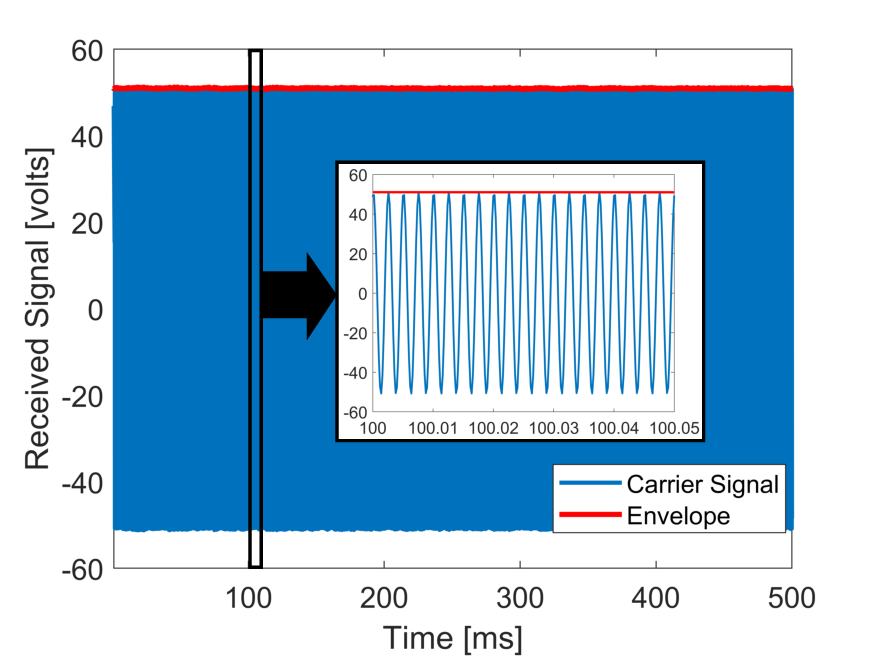}%
\label{fig_first_case}}
\hfil
\subfloat[]{\includegraphics[width=0.3\linewidth]{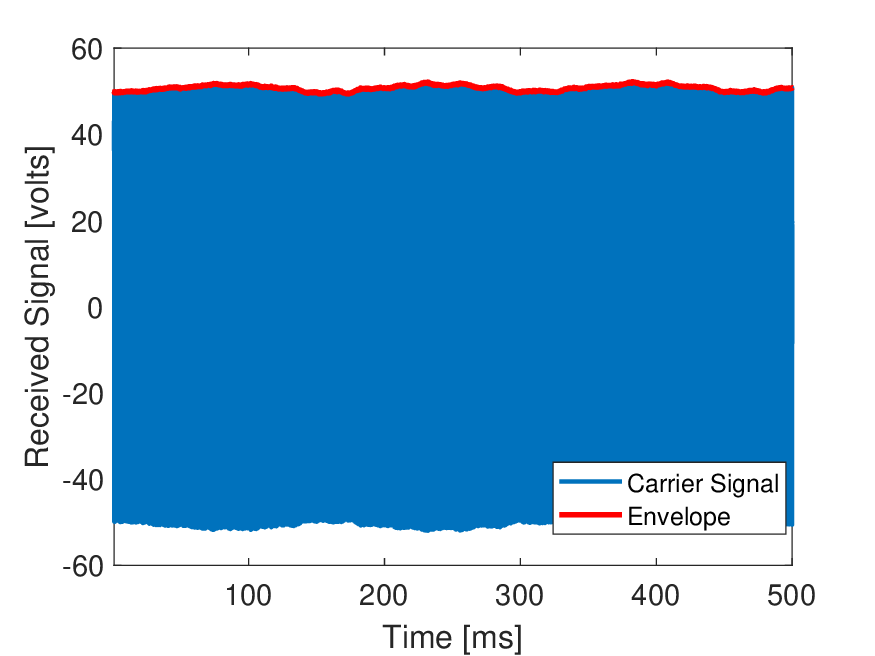}%
\label{fig_second_case}}
\hfil
\subfloat[]{\includegraphics[width=0.3\linewidth]{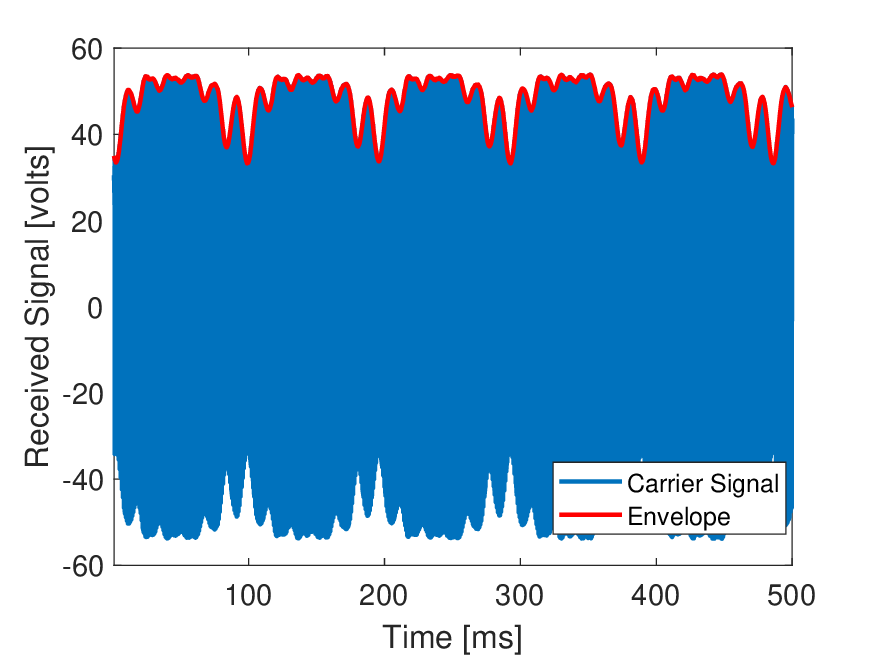}%
\label{fig_third_case}}
\caption{Received signal showing amplitude under three different jitter levels (a) Baseline - no intentional motion (b) low jitter (c) high jitter.}
\label{fig:high-jitter-envelope}
\end{figure*}

\subsection{Extraction of Gain Distribution}
\label{sect:extr-gain-distr}
A real-world communication channel is impacted by multiple factors. To isolate the effects of misalignment gain, other parameters such as antenna gains and propagation effects (e.g. absorption) must be normalized out of the analysis. Consider the typical time domain model used in analytical jitter models for a received signal with amplitude $y$:

\begin{equation}
y = hs+n
\label{eqn:signal-model-basic}
\end{equation}

\noindent where $s$ is the transmitted signal and $n$ represents additive white Gaussian noise (AWGN). The channel coefficient $h$ encompasses multiple effects including the misalignment gain $h_m$. Here we consider several standard parameters. These include free space path loss ($h_{pl}$) and atmospheric absorption ($h_a$). Note that $h_m$ is the only random variable in the channel coefficient. Multipath fading is assumed to be negligible. This is justified by previous experimental work, which has found the THz band to be a non-rich multipath environment, even in indoor settings with many potential scatterers \cite{papasotiriou_experimentally_2021}. The transmit and receive antenna gains are represented by $G_t$ and $G_r$, respectively. Therefore, the full signal model is:

\begin{equation}
y = h_{pl}h_ah_m\sqrt{G_tG_r}s+n = Ch_ms+n,
\label{eqn:signal-model-all-coefficients}
\end{equation}

\noindent where the constant parameters of our experiment have been combined into a single parameter $C$ for clarity, and thermal noise in the receiver has been assumed dominant over other noise sources. We wish to measure the distribution of $h_m$ by observing the measured signal $y$. Recall that the transmitted signal is a simple carrier tone $s = V_{tx}\sin{(2\pi f_ct)}$ where $V_{tx}$ is the transmitted signal amplitude in volts. After downconversion and filtering, the received baseband signal is:

\begin{equation}
y = V_{rx}h_m\sin{(2\pi f_{IF}t)}+n,
\label{eqn:signal-model-baseband}
\end{equation}

\noindent where $V_{rx} = CV_{tx}$ is the received signal amplitude accounting for all static gains and losses in the system and $f_{IF} = f_c-f_{LO}$. The IF frequency is selected by tuning the local oscillator frequency $f_{LO}$, which is supplied by a stable performance signal generator (PSG). 

Examining the envelope of $y$ is key to quantifying the effects of jitter. Under perfect alignment (zero jitter) conditions, $h_m = h_{m_{max}}$ will be constant, and so $y$ will have a fixed amplitude envelope $\varepsilon_{0} = V_{rx}h_{m_{max}}$. A measured example of this can be seen in Figure \ref{fig_first_case}, where the tone envelope (red line) is nearly constant and averages to a  voltage of $\varepsilon_{0} = 51.2$~mV. When jitter increases, $h_m$ becomes a random variable and the carrier tone amplitude fluctuates with the changes in $h_m$. Examples of this can be seen in Figures \ref{fig_second_case} and \ref{fig_third_case}. We are not  interested in the exact values of gain, path loss, or any of the other constant channel coefficients which set $V_{rx}$. Instead, we seek to isolate the jitter-induced fluctuations and histogram them to measure the distribution of $h_m$. The carrier envelope under jitter will be given by $\varepsilon = V_{rx}h_m+n$, and $h_m$ can be  isolated and normalized using the baseline measurement:

\begin{equation}\Bar
{\varepsilon} =  \frac{\varepsilon}{\varepsilon_0} = \frac{h_m}{h_{m_{max}}}+\frac{n}{V_{rx}h_{m_{max}}} = \Bar{h}_m+\Bar{n}.
\label{eqn:signal-norm-env}
\end{equation}

\noindent This normalized misalignment gain is  conveniently defined on the range $0 \leq \Bar h_m \leq 1$, making it easily compared to the analytical models, which are normalized by the ideal fraction of collected power $A_0$ (see Eqs.~\ref{eqn:h_m-formula} and \ref{eqn:w_eq}). Note that $\Bar{n}$ is merely the measured noise distribution rescaled by the baseline measurement value. Since the noise contribution is small relative to jitter, $\Bar{\varepsilon} \approx \Bar{h}_m$, and the histogrammed envelope values may be compared against the analytical distribution as described below.

\section{Results and Comparison with Literature}
In this section, we review the most commonly used model for the misalignment gain under jitter conditions before comparing its prediction with the measured results. 
\subsection{FSO-based model}

One of the earliest analytical models for THz jitter was derived from the FSO literature  \cite{abdalla_performance_2023,boulogeorgos_analytical_2019,boulogeorgos_outage_2020,boulogeorgos_joint_2022}. These works derive the distribution of misalignment gain as:
\begin{equation}
f_{h_m}(x) = \frac{\gamma^2}{A_0^{\gamma^2}}{x}^{\gamma^2-1}, \;\;\;\; 0 \leq x \leq A_0.
\label{eqn:h_m-formula}
\end{equation}

\begin{figure}[ht]
\centering
\includegraphics[width=1.0\linewidth]{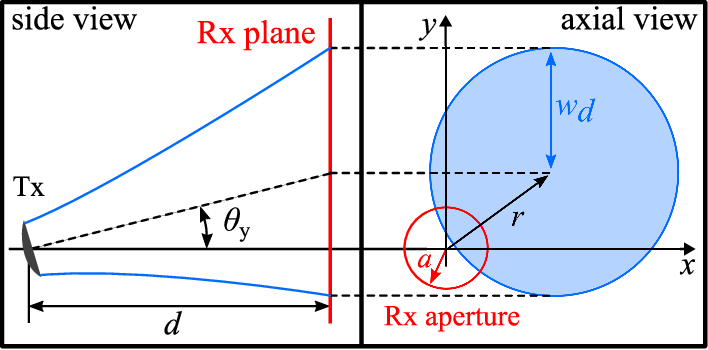}
\caption{Beam wander at the receiver. A Gaussian beam from the Tx arrives at the Rx with radius $w_d$.  The Rx radius is $a$. Azimuth and elevation deviations ($\theta_x, \theta_y)$ cause the beam center to wander in the Rx $xy$-plane with a displacement vector $r$. Only the elevation deviation $\theta_y$ is shown on the left.}
\label{fig:beamFootprint}
\end{figure}

\noindent This model is derived from the system illustrated in Fig.~\ref{fig:beamFootprint}. A Gaussian beam is transmitted over the distance $d$, arriving at the receiver with radius $w_d$. Jitter causes random shifts of the antenna in both the azimuth ($\theta_x$) and elevation ($\theta_y$) directions. These translate into shifts $r_x$ and $r_y$ in the $xy$-plane of the receiver, with net displacement $r = [r_x^2+r_y^2]^{1/2}$. Eq.~\ref{eqn:h_m-formula} is derived by assuming a Gaussian distribution of $r_x$ and $r_y$.

The parameter $A_0 = \mathrm{erf}^2(u)$ is the fraction of transmitted power collected by the receiver under perfect alignment, while $\gamma$ quantifies the jitter deviation versus beamwidth. The value of $\gamma$ sets the shape of the distribution, and is derived starting from the equivalent beam width at a receiver of radius $a$: 

\begin{equation}
w_{eq}^2 = w_d^2\frac{\sqrt{\pi}\mathrm{erf}(u)}{2u\:\mathrm{exp}(-u^2)}, \;\;\;\; u = \frac{\sqrt{\pi}a}{\sqrt{2}w_d}.
\label{eqn:w_eq}
\end{equation}

\noindent Then the ratio $\gamma$ is defined as

\begin{equation}
\gamma = \frac{w_{eq}}{2\sigma_r}.
\label{eqn:gamma}
\end{equation}

\noindent The parameter $\gamma$ is listed in terms of linear variance, but the total angular variance is more informative. Under the small angle approximation we may take $\theta = \sqrt{\theta_x^2 + \theta_y^2}$. This net angular deviation is related to the net linear deviation by:

\begin{equation}
\sigma_r = d\,\mathrm{tan}(\sigma_{\theta})\quad \leftrightarrow\quad \sigma_{\theta} = \tan^{-1}(\sigma_r / d),
\label{eqn:angle-std}
\end{equation}

\noindent where $\sigma_{\theta}$ is the random angular deviation.

To our knowledge, previous work has not considered the mean and variance of this distribution, preferring to directly derive the average BER or outage probability instead. However our novel experimental verification requires fitting of the measured data to the model, and these are useful metrics to compare. Analytical mean and variance are derived by inserting Eq.~\ref{eqn:h_m-formula} into the standard mean and variance definitions,

\begin{equation}
E[X] = \int_{-\infty}^{\infty}{x}f_{h_m}(x)dx 
= \frac{\gamma^2A_0}{\gamma^2+1} = \mu,
\label{eqn:mean}
\end{equation}

\begin{equation}
\mathrm{var}[X] = \int_{-\infty}^{\infty}\hspace{-0.5em}(x - E[X])^2f_{h_m}(x)\,dx = \frac{\gamma^2A_0^2}{\gamma^2 + 2} - \mu.
\label{eqn:var}
\end{equation}

\begin{figure}[h]
\centering
\includegraphics[width=1.0\linewidth]{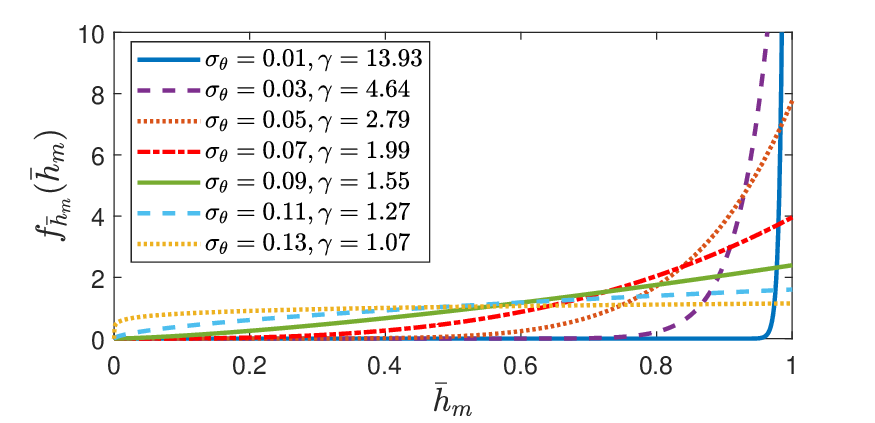}
\caption{Analytical model from Eq.~\ref{eqn:h_m-formula} under increasing jitter severity. The model is evaluated using the dimensions of the experimental setup. A receiver radius $a = 152.4$~mm, Gaussian transmitted beam with starting waist 152.4~mm, and transmission distance $d = 341$~m are assumed.}
\label{fig:PDF-examples}
\end{figure}

It is worthwhile to explore the behavior of this model under varying jitter conditions. The analytical model presented in Eq.~\ref{eqn:h_m-formula} is plotted in Fig.~\ref{fig:PDF-examples} using the physical dimensions of the experimental setup. A Gaussian beam of starting waist $152.4$~mm is modeled over a propagation distance of $d = 341$~m at 130~GHz, resulting in $w_d= 1.651$~m and collected by a receiver of radius $a = 152.4$~mm. Finally, the resulting analytical misalignment gain is normalized as $\bar h_m = h_m/A_0$ and plotted. The angular deviation is increased from small to large values. The specific values of $\sigma_{\theta}$ are chosen to produce illustrative values of the ratio $\gamma$. Large values ($\gamma > 10$) indicate jitter-induced beam wander that is small compared to $w_{eq}$, and these produce a delta-like distribution around the peak collected power ($\bar{h}_m=1$). When the  beam wander becomes comparable to $w_{eq}$ ($\gamma \approx 1$), the PDF approaches a flat distribution. The trend of these curves from low to high angular deviation should be noted for comparison to the measured data.  

\vspace{-0.1in}
\subsection{Measured Data and Analytical Fits}
\begin{figure*}[!ht]
\centering
\subfloat[]{\includegraphics[width=3.3in]{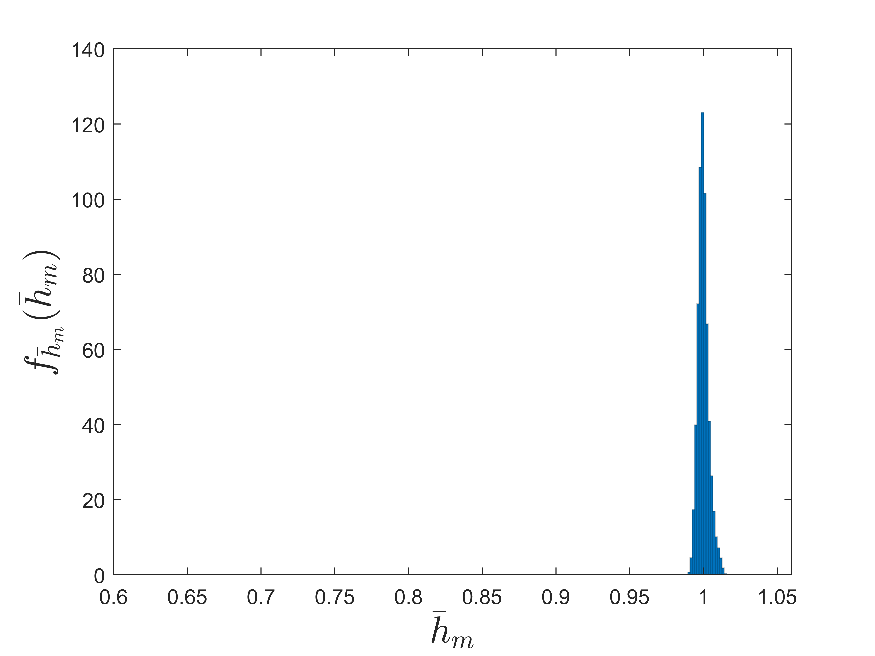}%
\label{baselinePDF}}
\hfil\vspace{-0.05in}
\subfloat[]{\includegraphics[width=3.3in]{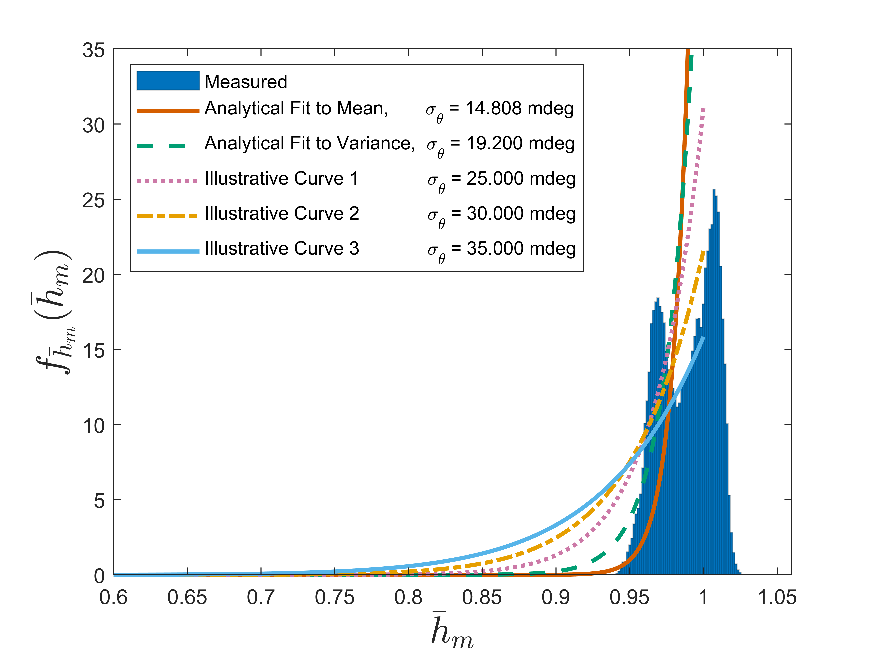}%
\label{jitter1PDF}}
\hfil\vspace{-0.05in}
\subfloat[]{\includegraphics[width=3.3in]{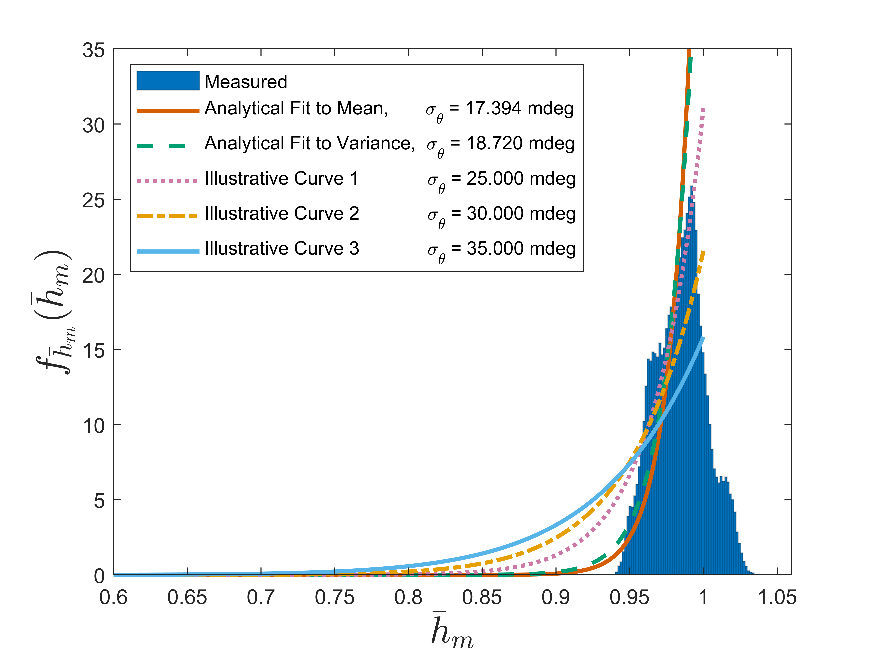}%
\label{jitter2PDF}}
\hfil
\subfloat[]{\includegraphics[width=3.3in]{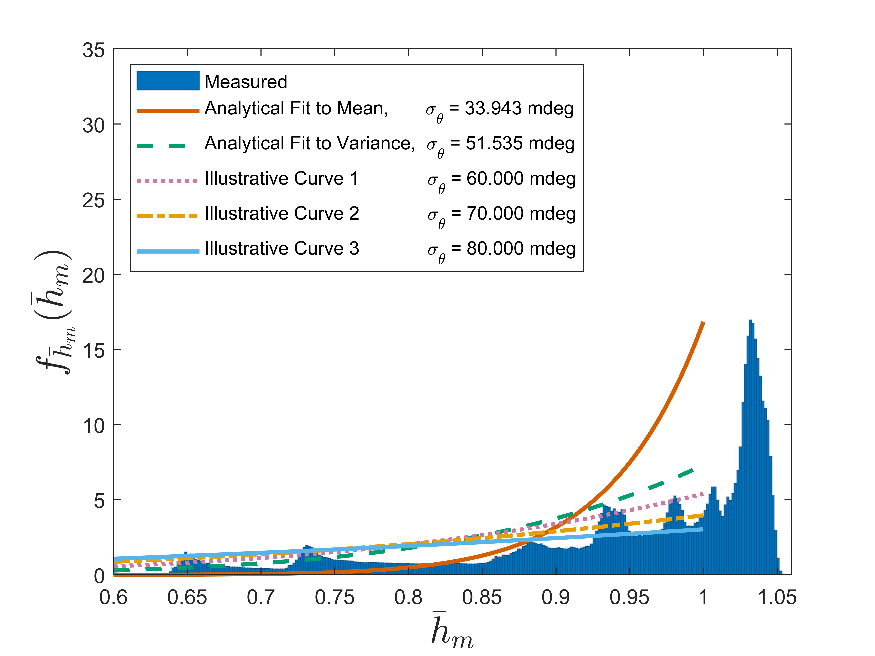}%
\label{jitter3PDF}}
\caption{Measured and normalized distribution of misalignment gain under increasing jitter (blue bins) and estimated analytical fits (red curves) (a) Baseline - no intentional jitter (b) low jitter (c) medium jitter (d) high jitter. The analytical model results have been overlaid on each measurement for various $\sigma_{theta}$. These curves are chosen to match the mean, variance, and highest peak of the measured PDFs.}\vspace{-0.1in}
\label{fig:jitterPDFs}
\end{figure*}

The procedure outlined in Section \ref{sect:extr-gain-distr} was used to generate histograms of the measured signal under each jitter condition. These are shown in Fig.~\ref{fig:jitterPDFs} (blue histograms). Fig.~\ref{baselinePDF} shows the baseline distribution with no jitter for reference, while the remaining subfigures illustrate how the distribution changes under increasing jitter. We note that these distributions are multi-modal, and that the number of peaks increases with jitter.  The mean and variance of these measured distributions are listed in Table \ref{table:jitter_stats}.

\begin{table}[!h]
\renewcommand{\arraystretch}{1.1}
\caption{Statistical Parameters of the Measured Data}
\label{table:jitter_stats}
\centering
\begin{tabular}{|c||c|c|}
\hline
Category & Mean & Variance\\
\hline
Baseline & 1.0 & 0.000016\\
Low Jitter & 0.988825 & 0.000335\\
Moderate Jitter & 0.984646 & 0.000304\\
High Jitter & 0.943949 & 0.011380\\
\hline
\end{tabular}
\end{table}

It should be immediately clear that the measured data does not match the expected forms shown in Fig.~\ref{fig:PDF-examples}. To further illustrate this, analytical PDFs were generated using Eqs.~\ref{eqn:h_m-formula}-\ref{eqn:w_eq} and overlayed with the measured data in Fig.~\ref{fig:jitterPDFs}. Values of $\gamma$ were chosen to fit statistical metrics of the measured data, including mean and variance. Then Eqs.~\ref{eqn:gamma}-\ref{eqn:angle-std} were used to determine the corresponding angular deviation $\sigma_{\theta}$. These curves, marked as ``Analytical Fit to Mean'' and ``Analytical Fit to Variance'' give an estimate of the net angular variation that would most closely match the measured distributions under the assumptions of Eq.~\ref{eqn:h_m-formula}. It was found that no analytical PDF can be generated using this model which matches both the mean and variance of the measured distribution. For illustrative purposes, several additional curves were plotted at increasing values of $\sigma_{\theta}$. These ``Illustrative Curves'' demonstrate that there is no choice of $\gamma$ (and by extension, $\sigma_{\theta}$) that can satisfactorily match the measured data. It is impossible to even approximate the shape of the measured distributions using Eq.~\ref{eqn:h_m-formula}. This is the key finding of this work, and it has important implications for bit error rate (BER) and outage probability calculations since these rely heavily on the channel coefficient distribution. 

\subsection{Discussion of the Measured Distributions}

The distributions in Fig.~\ref{fig:jitterPDFs} reveal important and unexpected behavior compared to the analytical models. We first address the upper tails of the distributions. The baseline measurement shown in Fig.~\ref{baselinePDF} is nearly Gaussian, with a mean value of 1.0. AWGN cannot be eliminated, and therefore even the normalized baseline distribution has small tails that extend slightly above 1.0. The distributions under jitter also show upper tails that exceed 1.0. This is partly due to the presence of thermal noise, as in the baseline measurement. However, these distributions extend further beyond 1.0 than what would be expected from only AWGN, based on the baseline measurement. This is not due to the involvement of multiple lobes of the antenna pattern.  The antenna angular motion was always significantly less than the $0.7^{\circ}$ first null beamwidth, meaning only the main lobe of the antenna pattern was involved.  Also, the obstruction-free LOS path is expected to be a non-rich multipath environment based on the findings in \cite{papasotiriou_experimentally_2021} and our own qualitative observations. Therefore there are two likely explanations for the tails above 1.0. One is initial alignment error. If a slight pointing error was present during the baseline measurement, then the normalization value $\epsilon_0$ would be set lower than the peak signal. Jitter-induced motion of the antenna would sweep through higher gain values, resulting in higher received values. Alignment of the system is extremely sensitive and involves a large number of adjustments.  This slight initial error is easily made given the millidegree sensitivity required. Additionally, the Cassegrain design is vulnerable to changes in the alignment of the main- and sub-reflectors caused by transient warping, which may slightly improve the focus of the antenna.

The shape of the distributions under jitter remains the most important result in this letter. At low jitter levels, Fig.~\ref{jitter1PDF} reveals a bimodal distribution. This alone is a significant deviation from the predicted range of possible behavior illustrated in Fig.~\ref{fig:PDF-examples}. However, the measured distributions become even more complex as jitter amplitude increases. The distribution becomes multi-modal, ending in the extreme spreading shown in Fig.~\ref{jitter3PDF} with 7 clearly visible modes. This multi-modal nature of the misalignment gain is not predicted by existing models. Furthermore, it is evident from the attempted analytical fits in Fig.~\ref{fig:jitterPDFs} that the analytical model \textit{cannot} match these distributions under any parameter combination. Even neglecting the multiple modes, the analytical model does not match the general shape of the measured distributions. Using the analytical model to predict performance metrics such as outage probability or BER in this system could produce highly inaccurate results. 

Some intuition for this drastic difference can be developed by considering the nature of the antenna motion. The existing literature follows the example of FSO jitter models and assumes a Gaussian distribution of the angular variance in azimuth and elevation. This is likely an accurate assumption for long distance FSO links, where micro-radian precision is often required and active beam steering is used. However, this work has demonstrated that, in at least some situations, this assumption may be incorrect for THz communication systems. In this experiment, a driven oscillation was introduced by a mechanical vibration source. This produces a \emph{periodic} swinging of the receive antenna rather than the assumed Gaussian-distributed motion. A sinusoidally oscillating antenna (or any driven mechanical structure with resonant vibrational modes) must decelerate at the end of travel and then accelerate again in the opposite direction. This causes the antenna to dwell at certain positions in the motion profile. Multiple mechanical vibrational modes may be excited in the antenna as jitter intensity increases, resulting in more complex motion with multiple direction reversals. This effect can be directly observed in the measured time-domain waveforms of Fig.~\ref{fig:high-jitter-envelope}.  Numerous peaks and valleys can be seen in the envelope of this time-domain signal.  This also explains the many modes seen in Fig.~\ref{jitter3PDF}. In fact, the number of unique peaks and valleys in Fig.~\ref{fig:high-jitter-envelope} corresponds to the number of modes in Fig.~\ref{jitter3PDF}, as would be expected.  We further note that no pre-meditated effort was made to design the antennas or jitter driving apparatus to exhibit mechanical jitter resonances.  The appearance of multi-modal jitter behavior was an unexpected result arising from simply applying a vibration source to an existing (and generally stable) antenna mount. This suggests that this effect will appear routinely in real-world THz backhaul links, and must be accurately addressed.  Analytical models are currently in development by the authors to capture these new effects. 

\section{Conclusion and Future Directions}
We have experimentally measured the effects of jitter on a real-world communication link placed in a realistic backhaul environment. A mechanical disturbance was applied to the system, such as might be caused by nearby machinery, wind loading, and other sources of mechanical excitation. The distribution of misalignment gain in this environment was found to deviate significantly from the usual analytical predictions derived from FSO work. Instead, complex and dynamic multi-modal distributions arose, stemming from the periodic nature of mechanical vibrations in the antenna's structure. 

This work shows that the difference between measured and analytical models has important implications for the accurate prediction of jitter effects and the design of link budgets for THz wireless backhaul links. It also shows the importance of understanding any mechanical resonances that may exist in antenna support structures.  On the same note, this work implies that stable and damped mechanical structures could significantly alleviate some of these  concerns associated with jitter. Future work will explore the mathematical foundations of this jitter in greater detail, and will investigate its quantitative effects on bit error rate (BER) as well as potential applications for novel sensing/diagnostics.

\section*{Acknowledgment}
We thank Drs. Shuai Nie and Shahriar Shahabuddin for their insightful comments on this work.

\bibliographystyle{IEEEtran}
\bibliography{refs}

\end{document}